# Relating $U(1)$ monopole configurations to $SU(2)$ saddle-point configurations

Chulwoo Jung [a] [*]

[a] Department of Physics, Columbia University, New York, NY 10027, USA

We have studied field configurations of the 3-dimensional Georgi-Glashow model which interpolate between the $U(1)$ and the $SU(2)$ limits. In the intermediate region, these configurations contain 't-Hooft–Polyakov monopoles. We use cooling and extremization to find these configurations and investigate their evolution as we adiabatically move towards the $U(1)$ and the $SU(2)$ limits. We also evolve an $SU(2)$ saddle point solution towards the $U(1)$ limit to see the relation between the unstable solutions in the $SU(2)$ theory and the stable ones in the $U(1)$ theory.

## 1. Introduction

This note is an update in our ongoing investigation of the role of semiclassical field configurations in the non-perturbative physics of 3-dimensional non-Abelian gauge theories. Previously, cooling [1,2] has been widely used to investigate the underlying long-range structure of wildly fluctuating gauge fields. Although cooling effectively brings configurations to relatively smooth ones, we can apply only moderate cooling to pure $SU(2)_3$ configurations because excessive cooling will eventually bring a configuration to vacuum (there are no nontrivial stable solutions of the equations of motion in $SU(2)_3$).

In contrast, extremization (deterministic reduction of $\hat{S}$ [3,4]) will bring configurations to the nearest, possibly unstable, solution of the equations of motion. We have expanded our earlier $SU(2)_3$ extremization algorithm to include the Georgi-Glashow model.

We use a discretized version of the Georgi-Glashow Lagrangian with the radially frozen approximation ($\vec{\phi}_n \cdot \vec{\phi}_n = 1$, $\vec{\phi}_n$ is a real 3-vector).

$$S(\beta_g, \beta_h) = \beta_g S_g + \beta_h S_h \equiv \frac{1}{2}\beta_g \sum_P tr(U_P)$$
$$+ \frac{1}{2}\beta_h \sum_{n\mu} tr(\vec{\phi}_n \cdot \vec{\sigma} U_{n\mu} \vec{\phi}_{n+\hat{\mu}} \cdot \vec{\sigma} U_{n\mu}^\dagger)$$

[*] This work was done in collaboration with Robert D. Mawhinney. It was partially supported by the US Department of Energy and the Pittsburgh Supercomputing Center.

This model describes a $U(1)$ theory as $\beta_h \to \infty$ and the pure $SU(2)$ theory as $\beta_h \to 0$. In the intermediate region, we have nontrivial solutions which are t'Hooft-Polyakov monopoles. The condensation of these monopole-antimonopole ($M$-$\bar{M}$) pairs is known to be responsible for confinement in this phase [5]. Studies showed no qualitative change in physical quantities such as the Creutz ratios when moving between the $U(1)$ and the $SU(2)$ limits [6].

In last year's proceedings [3], we presented a class of pure $SU(2)$ saddle point solutions which we will refer to as $Z(2)$ saddles. These saddles were made by extremizing $M$-$\bar{M}$ pairs in an arbitrary $U(1)$ subgroup. We also have stable $U(1)$ $M$-$\bar{M}$ solutions available. We use these two solutions – a $U(1)$ $M$-$\bar{M}$ pair and $Z(2)$ saddle in the $SU(2)$ limit – as starting configurations and evolve the solutions towards the other limits keeping $\hat{S}$ as small as possible. Along the way, we measured Creutz ratios and the eigenvalues of harmonic fluctuations around the final configurations [3].

## 2. Evolution of configurations

We started by cooling a $M$-$\bar{M}$ pair on a $16^3$ lattice separated by 6 lattice spacings at $\beta_g = 3.0$, $\beta_h = 3.0$ to completion. This was evolved, using the parameter set obtained by Duncan and Mawhinney [6], towards the $SU(2)$ limit using cooling to allow their non-Abelian cores to exceed one lattice spacing. (If we use extremization instead, the configuration will stay in the $U(1)$ subgroup.) At $\beta_g = 6.5$, $\beta_h = 1.0$, the configuration becomes unstable under cooling. We then used this as the starting configuration and extremized it, changing parameters to evolve the configuration both to the $U(1)$ and the $SU(2)$ limits.

As you can see in Table 1, $\hat{S}$ is quite small in the $U(1)$ and the $SU(2)$ limits, but remains relatively high in the region $0.0 < \beta_h \leq 1.25$. As we approach the $U(1)$ limit, the algorithm quickly finds a pure $U(1)$ solution with only positive eigenvalues. Table 3 lists eigenvalues from the Lanczos algorithm, where zero eigenvalues associated with gauge transformations are not shown. In the $SU(2)$ limit, we observed $S_g$ decreasing to zero. The action in the Higgs sector ($S_h$) increases but eventually becomes irrelevant as $\beta_h \to 0$.

Figure 1 shows the behavior of the Creutz ratios. We can see the Creutz ratios rising at small distances and falling at large distances because there is only one $M$-$\bar{M}$ pair. If we have monopole condensation, we will see a plateau at large distances. In the $SU(2)$ limit, the Creutz ratio retains its shape but the amplitude decreases to zero.

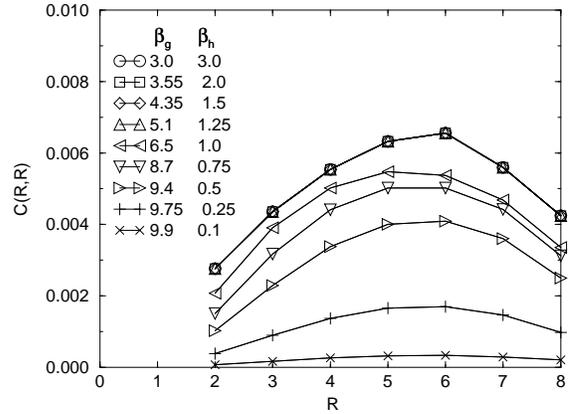

Figure 1. Creutz ratios for the configurations evolved from 't-Hooft–Polyakov $M$-$\bar{M}$ pair.

| $\beta_g$ | $\beta_h$ | $S/\beta_g$ | $\hat{S}/\beta_g^2$ | $S_g$ | $S_h$ |
|---|---|---|---|---|---|
| 6.5 | 1.0 | 6.486 | 8.05e-4 | 3.74 | 1.79e+1 |
| 5.1 | 1.25 | 8.966 | 3.41e-13 | 8.97 | 5.12e-11 |
| 4.35 | 1.5 | 8.966 | 1.40e-17 | 8.97 | 4.99e-11 |
| 3.55 | 2.0 | 8.966 | 6.65e-18 | 8.97 | 5.07e-11 |
| 3.0 | 3.0 | 8.966 | 3.36e-18 | 8.97 | 5.03e-11 |
| 6.5 | 1.0 | 6.486 | 8.05e-4 | 3.74 | 1.79e+1 |
| 8.7 | 0.75 | 4.9650 | 8.77e-3 | 2.45 | 2.90e+1 |
| 9.4 | 0.5 | 3.825 | 1.58e-2 | 1.58 | 4.21e+1 |
| 9.75 | 0.25 | 2.354 | 1.23e-2 | 5.85e-1 | 6.91e+1 |
| 9.9 | 0.1 | 1.088 | 4.91e-3 | 1.11e-1 | 9.67e+1 |
| 9.95 | 0.05 | 0.569 | 1.60e-3 | 2.98e-2 | 1.07e+2 |
| 9.97 | 0.03 | 0.349 | 7.58e-4 | 1.25e-2 | 1.12e+2 |

Table 1
The action and Hessian ($\hat{S}$) for a t'Hooft-Polyakov $M$-$\bar{M}$ cooled to completion at $\beta_g = 6.5$, $\beta_h = 1.0$ and then evolved to the $U(1)$ and $SU(2)$ limits.

Starting from a single $Z(2)$ saddle in the unbroken phase on a $12^3$ lattice, we gauge transformed to maximal Abelian gauge and added a Higgs field pointing in the $\sigma_3$ direction. This minimizes the possible disturbance introduced by the Higgs field. We then moved adiabatically towards the $U(1)$ limit.

In the $U(1)$ limit, our $Z(2)$ solutions become pure $U(1)$. But, contrary to our expectation, this $U(1)$ solution was not a stable one as you can see from the negative eigenvalues in Table 3. Figure 2 shows the behavior of the Creutz ratios. Near the $SU(2)$ limit, the Creutz ratios rise at large distances, a characteristic shared by extremized Monte Carlo lattices [4]. As we approach the $U(1)$ limit, the Creutz ratio curve changes its shape and becomes similar to that of a monopole-antimonopole pair. We should point out that this solution contains a $U(1)$ $M$-$\bar{M}$ pair, but since it has unstable modes, it differs from the simple $U(1)$ $M$-$\bar{M}$ pair discussed in Table 1.

## 3. Discussion

There were a few surprises along our course of investigation. Firstly, in the intermediate region, we could not make $\hat{S}$ arbitrarily small. At this point, we are not sure if this comes from some fundamental aspects of the model or a drastic decrease in our algorithm's efficiency. (We have in-vestigated whether this pathology is coming from our particular choice of the discretization of Lagrangian. We have relaxed the radially frozen approximation and added $\phi^4$ and $\phi^6$ terms and the result remains the same.)

| $\beta_g$ | $\beta_h$ | $S/\beta_g$ | $\hat{S}/\beta_g^2$ | $S_g$ | $S_h$ |
|---|---|---|---|---|---|
| 10.0 | 0.0 | 2.299 | 4.10e-17 | 2.30 | |
| 9.9 | 0.1 | 2.832 | 1.78e-5 | 2.33 | 4.88e+1 |
| 9.75 | 0.25 | 3.516 | 2.78e-5 | 2.49 | 4.01e+1 |
| 9.4 | 0.5 | 4.480 | 2.17e-5 | 2.83 | 3.08e+1 |
| 8.7 | 0.75 | 5.383 | 2.32e-5 | 3.28 | 2.43e+1 |
| 6.5 | 1.0 | 6.745 | 2.00e-6 | 4.16 | 1.68e+1 |
| 5.1 | 1.25 | 7.999 | 3.90e-8 | 5.25 | 1.12e+1 |
| 4.35 | 1.5 | 8.918 | 6.07e-8 | 6.42 | 7.46 |
| 3.55 | 2.0 | 9.950 | 4.64e-8 | 8.56 | 2.47 |
| 3.0 | 3.0 | 10.145 | 1.19e-8 | 1.01e+1 | 3.82e-8 |

Table 2
The action and Hessian for configurations derived from a $Z(2)$ saddle.

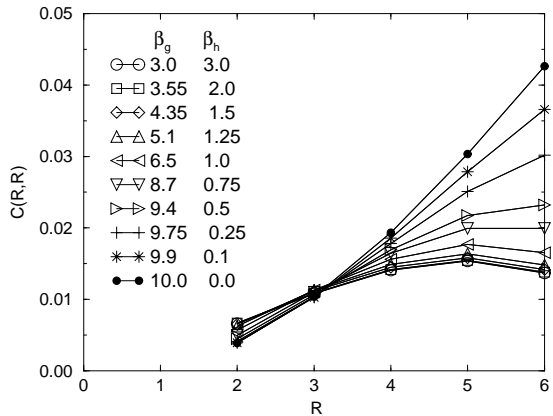

Figure 2. Creutz ratios for configurations generated from $Z(2)$ saddle.

Secondly, the t'Hooft-Polyakov $M$-$\bar{M}$ pair solution we created seems to evolve to the vacuum when we approached the pure $SU(2)$ limit. We need more configurations – possibly made from Monte Carlo lattices – to generalize this result. Also, we need a better understanding of the difference between the simple $M$-$\bar{M}$ pair in the $U(1)$ limit and the saddle point solution we got from our $Z(2)$ saddle.

| $Z(2)$ saddle | | $U(1)$ $M$-$\bar{M}$ |
|---|---|---|
| $\beta_g = 10.0$ | $\beta_g = 3.0$ | $\beta_g = 3.0$ |
| $\beta_h = 0.0$ | $\beta_h = 3.0$ | $\beta_h = 3.0$ |
| -1.7105e+00 | -2.9161e+01 | |
| -5.6828e-01 | -2.9161e+01 | 4.5471e-01 |
| -5.6828e-01 | -2.9144e+01 | 4.5566e-01 |
| -3.8832e-01 | -2.9144e+01 | 4.5592e-01 |
| -3.8832e-01 | -1.9543e+01 | 4.5666e-01 |
| -3.8832e-01 | -1.9541e+01 | 4.5669e-01 |
| | -3.6402e-01 | 4.5670e-01 |
| 9.8356e-01 | -1.1682e-01 | 4.5671e-01 |
| 9.8356e-01 | | 9.0869e-01 |
| 1.5262e+00 | 3.1096e-03 | 9.0871e-01 |
| 1.7331e+00 | 3.1190e-03 | 9.1029e-01 |

Table 3
The smallest eigenvalues for harmonic fluctuations around our saddle point solutions from the Lanczos algorithm[7]. We have not listed the zero eigenvalues associated with gauge invariance.

We are planning to generate ensembles of extremized lattices (both $U(1)$ and $SU(2)$) and evolve them to the other limit to investigate the behavior of the string tension and unstable modes further. We are also pursuing finding saddle point solutions in the intermediate region.